\documentclass{llncs}

\usepackage{graphicx, verbatim}

\begin{document}

\title{Frozen Footprints}

\author{Massimo Franceschet}

\institute{Department of Mathematics and Computer Science, University of Udine \\
           Via delle Scienze 206 -- 33100 Udine, Italy \\
           \texttt{massimo.franceschet@dimi.uniud.it}}

\maketitle

\begin{abstract}
Bibliometrics has the ambitious goal of measuring science. To this end, it exploits the way science is disseminated trough scientific publications and the resulting citation network of scientific papers. We survey the main historical contributions to the field, the most interesting bibliometric indicators, and the most popular bibliometric data sources. Moreover, we discuss  distributions commonly used to model bibliometric phenomena and give an overview of methods to build bibliometric maps of science.
\end{abstract}

\section{Introduction} \label{introduction}

Bibliometrics is a research method used in library and information science. It uses quantitative analysis and statistics in order to:

\begin{itemize}

\item determine the influence of single scholars or groups of them (e.g., research groups, institutions, countries) and that of single papers or groups of them (e.g., journals or entire research fields);

\item describe the relationships between authors, publications, journals, or research fields.

\end{itemize}

Bibliometrics has become a standard tool of science policy and research management in the last decades. Academic institutions increasingly rely on bibliometric analysis for making decisions regarding hiring, promotion, tenure, and funding of scholars; authors, librarians, and publishers may use citation indicators to evaluate journals and to select those of high impact; editors may choose reviewers on the basis of their bibliometric scores on a particular subject of interest; worldwide college and university rankings, e.g., THE-QS\footnote{\texttt{http://www.timeshighereducation.co.uk/}} and ARWU\footnote{\texttt{http://www.arwu.org/}}, which are partially based on bibliometric criteria, are often consulted by prospective students and their parents in the college and university admissions process. Today, bibliometrics is one of the rare truly interdisciplinary research fields, with important links with history of science, sociology, law, economics, management, theology, mathematics, statistics, physics, and computer science.

Citation analysts retrieve production and citation data from bibliometric data sources and compute performance indicators to measure the quality of research of an actor.  The bibliographic databases of the Institute for Scientific Information, today Thomson-Reuters, have been used for decades as a starting point and often as the only tools for locating citations and conducting citation analysis. Fierce competitors of the databases provided by Thomson-Reuters are Elsevier's Scopus and the freely accessible Google Scholar.

Actors under evaluation are typically individual scholars and journals, but bibliometric units can be composed of homogeneous groups of scholars or groups of journals at different levels of aggregation. Bibliometric criteria that characterize research quality are productivity, impact (or popularity), and prestige. Typically bibliometric indicators capture separately some of these criteria, hence the need of using in the evaluation process several orthogonal metrics capturing different aspects of research performance.

The outline of this manuscript is as follows. We first briefly review the main historical contributions to the field in Section \ref{history}. In Section \ref{citations} we discuss the controversial role of citations in bibliometrics. Section \ref{indicators} surveys the most interesting bibliometric indicators both at the individual and at the journal level, while Section \ref{databases} is devoted to the comparison of the most popular bibliometric data sources. Section \ref{distributions} investigates the probability distributions that underlie most phenomena in bibliometrics. In Section \ref{maps} we delve into the realm of bibliometric maps of science. Finally, Section \ref{quote} contains some of the best quotations about bibliometrics.

\section{Historical remarks} \label{history}

Bibliometric studies started long time ago. A remarkable early piece of work is \textit{Histoire des sciences et des savants depuis deux si\`{e}cles}. The author, Alphonse de Candolle, describes the scientific strength of nations and tries to find environmental factors for the scientific success of a nation \cite{C73}.

Derek John de Solla Price (1922 -- 1983), an historian of science and information scientist born from Philip Price, a tailor, and Fanny de Solla, a singer, is credited as the father of bibliometrics. In his book \textit{Little Science, Big Science}, he analyzed the recent system of science communication and laid the foundation of modern research evaluation techniques \cite{S63}.

The term \textit{bibliometrics} is introduced by Pritchard in 1969 \cite{P69}. Pritchard explains the term bibliometrics as:

\begin{quote}
\textit{the application of mathematical and statistical methods to books and other media of communication}
\end{quote}

\noindent At the same time, Nalimov and Mulchenko defined scientometrics as:

\begin{quote}
\textit{the application of those quantitative methods which are dealing with the analysis of science viewed as an information process}
\end{quote}

According to these definitions, scientometrics is restricted to science communication, whereas bibliometrics is designed to deal with more general information processes. Nowadays, the borderlines between the two specialities almost vanished and both terms are used almost as synonyms.

The statistical analysis of scientific literature began years before the term bibliometrics was coined. The main contributions are: Lotka's Law of scientific productivity, Bradford's Law of scatter, and Zipf's Law of word occurrence. In 1926, Alfred J. Lotka published a study on the frequency distribution of scientific productivity determined from a decennial index of Chemical Abstracts \cite{L26} (see Table \ref{Lotka}). Lotka concluded that:

\begin{quote}
\textit{In a given field, the number of authors making n contributions is about $1/n^2$ of those making one.}
\end{quote}

Lotka's Law means that few authors contribute most of the papers and many or most of them contribute few publications. For instance, in the original data of Lotka's study illustrated in Table \ref{Lotka}, the most prolific 1350 authors (21\% of the total) wrote more than half of the papers (6429 papers, 51\% of the total).

\begin{table}
\begin{center}
\begin{tabular}{|c|c|c|c|}
\hline
\textbf{papers} &	\textbf{observed authors} & \textbf{expected authors} &	\textbf{total papers}
\\ \hline
1 & 3991 & 3991 & 3991 \\ \hline
2 & 1059 & 998 & 2118 \\ \hline
3 & 493 & 443 & 1473 \\ \hline
4 & 287 & 249 & 1148 \\ \hline
5 & 184 & 160 & 920 \\ \hline
6 & 131 & 111 & 786 \\ \hline
7 & 85 & 81 & 595 \\ \hline
8 & 64 & 62 & 512 \\ \hline
9 & 65 & 49 & 585 \\ \hline
10 & 41 & 40 & 410 \\ \hline
\textbf{total} & \textbf{6400} & \textbf{6890} & \textbf{12538} \\ \hline
\end{tabular}
\end{center}
\caption{Original Lotka data and fit on the basis of Lotka distribution (up to 10 papers)}
\label{Lotka}
\end{table}

Eight years after Lotka's article appeared, Bradford published his study on the frequency distribution of papers over journals \cite{B34}. It states that:

\begin{quote}
\textit{If scientific journals are arranged in order of decreasing productivity on a given subject, they can be divided into groups of different sizes (number of journals) each containing the same number of papers relevant to the subject. The size of each group (except the first) is given by the size of the previous group multiplied by a constant.}
\end{quote}

Bradford formulated his law after studying a bibliography of geophysics (Table \ref{Bradford}). Journals can be divided in 3 groups of different sizes but containing about the same number of relevant papers: a core group, columns 1 and 2 of the table, containing 2 journals and 179 relevant papers, a second group, columns 3-6, containing 4 journals and 185 relevant papers, and a third group, columns 7-17, containing 11 journals and 186 relevant papers.

\begin{table}
\begin{center}
\begin{tabular}{|c|c|c||c|c|c|c||c|c|c|c|c|c|c|c|c|c|c|}
\hline
\textbf{journal} & 1 & 2 & 3 & 4 & 5 & 6 & 7 & 8 & 9 & 10 & 11 & 12 & 13 & 14 & 15 & 16 & 17 \\ \hline
\textbf{productivity} & 93 & 86 & 56 & 48 & 46 & 35 & 28 & 20 & 17 & 16 & 16 & 16 & 16 & 15 & 14 & 14 & 146 \\ \hline
\end{tabular}
\end{center}
\caption{Original Bradford data for subject geophysics. Bradford arranged journals in order of decreasing productivity on the subject and counted the number of papers in the journal that are relevant to geophysics.}
\label{Bradford}
\end{table}

What it means is that for each speciality there are few core journals for that field that contribute a relatively great amount of publications and many or most of the journals give few contributions. Hence, it is sufficient to identify the core journals for that field and look at them. Very rarely will researchers need to go outside that set. This may serve, for example, to guide librarians in choosing the core journals to stock in any given field. Bradford's Law also caused the discovery, which some did not expect, that a few journals like Nature and Science were core for all of hard science. The same pattern does not happen with the humanities or the social science -- possibly because objective truth is so much harder to establish there. The result of this is pressure on scientists to publish in the best journals, and pressure on universities to ensure access to that core set of journals.

Zipf formulated an interesting law in bibliometrics and quantitative linguistics that he derived from the study of word frequency in texts \cite{Z49}. Zipf's Law  states that:

\begin{quote}
\textit{In relatively lengthy texts, if words occurring within the text are listed in order of decreasing frequency, then the rank of a word on that list multiplied by its frequency will equal a constant, which depends on the analyzed text.}
\end{quote}

\begin{table}
\begin{center}
\begin{tabular}{|c|c|c|}
\hline
\textbf{rank} &	\textbf{frequency} & \textbf{rank * frequency}
\\ \hline
10 & 2653 & 26530 \\ \hline
20 & 1311 & 26220 \\ \hline
30 & 926 & 27780 \\ \hline
40 & 717 & 28680 \\ \hline
50 & 556 & 27800 \\ \hline
100 & 265 & 26500 \\ \hline
200 & 133 & 26600 \\ \hline
300 & 84 & 25200 \\ \hline
400 & 62 & 24800 \\ \hline
500 & 50 & 25000 \\ \hline
1000 & 26 & 26000 \\ \hline
2000 & 12 & 24000 \\ \hline
3000 & 8 & 24000 \\ \hline
4000 & 6 & 24000 \\ \hline
5000 & 5 & 25000 \\ \hline
10000 & 2 & 20000 \\ \hline
20000 & 1 & 20000 \\ \hline
29899 & 1 & 29899 \\ \hline
\end{tabular}
\end{center}
\caption{The distribution of words in Joyce's Ulysses}
\label{Zipf}
\end{table}

Zipf illustrated his law with an analysis of James Joyce's Ulysses (Table \ref{Zipf}). It means that only a few words are used very often, many or most are used rarely. Why natural language texts conform to Zipfian distribution has been a matter of some controversy. Zipf explains his law with the Principle of Least Effort, defined as follows:

\begin{quote}
\textit{Each individual will adopt a course of action that will involve the expenditure of the probably least average of his work.}
\end{quote}

According to Zipf, if the Principle of Least Effort works, the speaker (or writer) tends to minimize number and length of words (this he calls the Force of Unification), by overloading the same word with different meanings, while the hearer (or reader) calls for a diversification of words (this he calls the Force of Diversification), by assigning different meanings to different words. For communication to be effective, these opposite forces need to equilibrate, giving rise to the mentioned law of word occurrence.

Only in the beginning of the 1980's, with the fast development of computer science, bibliometrics could evolve into a distinct scientific discipline with a specific research profile and corresponding communication structures. \textit{Scientometrics}, the first international periodical specialized on bibliometric topics, started in 1979. The fact that bibliometric methods are already applied to the bibliometric field itself also indicates the rapid growth of the discipline.

\section{The role of citations} \label{citations}

A central question is: why bibliometric analysis of research performance? Peer review, that is, the evaluation made by expert peers, undoubtedly is an important procedure of quality judgment. In particular, the results of peer review judgment and those of bibliometric assessment are not completely independent variables. Indeed, peers take some bibliometric aspects into account in their judgment, for instance number of publications in the better journals.

But peer review and related expert-based judgments may have serious shortcomings. Subjectivity, i.e., dependence of the outcomes on the choice of individual committee members, is one of the major problems. Moreover, peer review is slow and expensive (at least in terms of hours of volunteer work devoted to refereeing). In particular, peer review methodology is practically unfeasible when the number of units to evaluate is consistent, e.g., all papers published by all members of a large department.

Bibliometric assessment of research performance is based on the following central assumptions \cite{vR06b}:

\begin{itemize}
\item scholars who have to say something important do publish their findings;

\item scholars refer in their own work to earlier work of other scholars to acknowledge intellectual debt and to witness the use of information.
\end{itemize}
In research evaluation, citations became a widely used measure of the \textit{impact} of scientific publication. Smith \cite{S81} stated that:

\begin{quote}
\textit{Citations are signposts left behind after information has been utilized.}
\end{quote}

\noindent while Cronin \cite{C81} defined citations as:

\begin{quote}
\textit{frozen footprints in the landscape of scholarly achievement which bear witness of the passage of ideas.}
\end{quote}

However, problems with citation analysis as a reliable instrument of measurement and evaluation have been acknowledged. Citations reflect both the needs and idiosyncrasies of the citer, including such factors as utility, quality, availability, advertising (self-citations), collaboration or comradeship (in-house citations), chauvinism, mentoring, personal sympathies and antipathies, competition, neglect, obliteration by incorporation, augmentation, flattery, convention, reference copying, reviewing, and secondary referencing~\cite{MM89}. As Seglen says (\cite{S92}, page 636), \textit{``while the sheer number of factors may help to achieve some statistical balance, we all know of scientists who are cited either much less (ourselves) or much more than they deserve on the basis of their scientific achievements''}.

Nevertheless, citation analysis has demonstrated its reliability and usefulness as a tool for ranking and evaluation scholars and their publications \cite{W88}. Furthermore, the robustness of citations as a method to evaluate impact is particularly witnessed by the adoption of a similar approach in several other fields far different from bibliometrics, including web pages connected by hyperlinks \cite{K99,PBMW99}, patents and corresponding citations \cite{N94}, published opinions of judges and their citations within and across opinion circuits \cite{LLS98}, and even sections of the Bible and the biblical citations they receive in religious texts \cite{MT05}.

\section{Bibliometric indicators} \label{indicators}

Assuming the central bibliometric assumptions mentioned in Section \ref{citations}, we may design quantitative indicators to assess research quality of an actor. But, what aspects characterize quality of research? Moreover, what are the actors under evaluation?

There is a general agreement that research quality is not characterized by a single element of performance. Van Raan \cite{vR06c} claims:

\begin{quote}
\textit{It is not wise to force the assessment of researchers or of research groups into just one measure, because it reinforces the opinion that scientific performance can be expressed simply by one note. Several indicators are necessary in order to illuminate different aspects of performance.}
\end{quote}

\noindent Moreover, Gl\"{a}nzel \cite{G06} adds:

\begin{quote}
\textit{the use of a single index crashes the multidimensional space of bibliometrics into one single dimension.}
\end{quote}

Two potential dangers of condensing down quality of research to a single metric are:

\begin{itemize}

\item a person may be damaged by the use of a simple index in a decision-making process if the index fails to capture important and different aspects of research performance;

\item scientists may respond to this by maximizing that particular metric to the detriment of doing more justifiable work.
\end{itemize}

Quality of research is therefore described by different aspects; the most important are:

\begin{itemize}

\item \textit{productivity}; this is the amount of scholarly works that are produced by the actor;

\item \textit{impact} (or popularity); this is the number of endorsements that the actors receives from other actors;

\item \textit{prestige}; this is the prestige of the works produced by the actor and that of the endorsing actors.

\end{itemize}

The actors under bibliometric evaluation may be different, however, the basic unit of evaluation is a single scholarly work, typically, a journal paper. This basic unit, a scholarly work, can be aggregated at different levels obtaining more complex units of evaluation. For instance, single scholars are evaluated in terms of the set of works they produced. Scholars are typically grouped into research groups, institutions, regions within nations, countries or even international regions. Moreover, scholarly works are typically aggregated into journals or conferences and research fields. The level of aggregation is:

\begin{itemize}
\item \textit{micro-level}, when individuals, research groups or single scholarly works are considered;
\item \textit{meso-level}, in the case of institutions or journals;
\item \textit{macro-level}, in the case of regions, countries or research fields.
\end{itemize}

Before delving into the realm of bibliometric indicators, it is important to understand that different scholarly disciplines can have very different publication and citation practices, including the absolute number of researchers, the average number of authors on each paper, the average number of citations in each paper, and the nature of results \cite{AM00}.
All these factors complicate the use of evaluation metrics across different disciplines. Nevertheless, interdisciplinary indicators have been proposed along the following line. In principle, it is possible to compute the mean (or median) number of citations per paper for an entire research discipline \cite{RFC08}. Hence, the actual number of citations for a publication can be compared with the expected number of citations for the field of the publication \cite{vR06b}.

\subsection{Bibliometric measures at the individual level}

The traditional bibliometric indexes used to evaluate the performance of individual scholars include:

\begin{itemize}
\item  the \textit{number of publications} produced by the scholar, possibly divided by the scholar's academic age;

\item the \textit{number of citations} that the publications produced by the scholar have received from other scholarly works, possibly divided by the number of publications.

\end{itemize}

A more interesting measure is the \textit{h index}. The h index of a scholar is the higher number of papers a scholar has that have each received at least that number of citations. For instance, my current h index computed with Google Scholar is 14, meaning that I am the author of 14 papers each of them cited at least 14 times. The rest of my papers are all cited a number of times that is less or equal to 14. The index was proposed by Hirsch, a physicist, in 2005 \cite{H05} and it has immediately found interest in the public \cite{B05,Ba07} and in the bibliometrics literature. In particular, it is currently computed by both Web of Science and Scopus.

The index is meant to capture both productivity and impact of a scholar in such a way that it is hard to increase it, as well as to rig it, over a certain threshold. It favors researchers who produce a continuous stream of influential papers over those who publish many quickly forgotten ones or a few blockbusters. Moreover, it is difficult to inflate the index, for instance with self-citations. Indeed, all self-citations to papers with less than h citations are irrelevant for the computation of the index, as are the self-citations to papers with many more than h citations.

Hirsch argues that the h index is preferable to other single-number criteria commonly used to evaluate scientific output of a researcher \cite{H05} and that it has more predictive power \cite{H07}. Hirsch suggests that, for a given researcher, h should increase approximately linearly with time, that is $h = m \cdot n$, where $n$ is the academic age in years and $m$ is the slope of the linear function. The parameter $m$ should provide a useful yardstick to compare scientists of different seniority. In particular:

\begin{itemize}
\item  a value of m around 1 characterizes a successful scientist;

\item a value of m around 2 characterizes outstanding scientists;

\item a value of m around 3 or higher characterizes truly unique individuals.

\end{itemize}

An additional advantage of the h index is that it is extremely simple and comprehensible. Moreover, it can be easily computed by sorting the published papers in decreasing order with respect to the number of received citations and scrolling down the list until the rank of the paper is greater than the number of citations that it has. The preceding rank equals the h index.

The h index has also been criticized; in particular, Gl\"{a}nzel \cite{G06} and Bornmann and Daniel \cite{BD07} describe opportunities and limitations of the h index.  The following are acknowledged limitations of the index:

\begin{enumerate}

\item it puts newcomers at a disadvantage since both publication output and citation rates will be relatively low;

\item it does not account for the number of authors in a paper;

\item it is discipline dependent;

\item it disadvantages small but highly-cited paper sets too strongly;

\item it allows scientists to rest on their laurels (``\textit{your papers do the job for you}'') since the index never decreases and it might increase even if no new papers are published.

\end{enumerate}

Many variations of the index have been proposed to correct the mentioned flaws:

\begin{itemize}

\item Hirsch proposes to solve problems number 1 and 5 by dividing the h index by the scientific age of the author \cite{H05};

\item to address problem number 2 and, partially, issue number 3, Batista et al.\ \cite{BCK06} suggest to adjust the original h index by dividing it by the mean number of researchers in the h publications of the Hirsch core;

\item Egghe \cite{E06} proposes the g-index to account for problem number 4. Given a set of articles ranked in decreasing order of the number of citations that they received, the g-index is the (unique) largest number such that the top g articles received (together) at least g$^2$ citations. Moreover, for the same problem, Jin \cite{J06} suggests to use the average number of citations received by articles in the Hirsch core (the set of articles that determine the h index value);

\item in order to address issue number 5, Katsaros et al.\ \cite{KMS07}  propose the contemporary h index. The contemporary h index adds an age-related weighting to each cited article, giving less weight to older articles.

\end{itemize}

In fact, all these variations did not attract much attention since they address a single issue without considering the others; hence the original version of the h index is still the most adopted.

\subsection{Bibliometric measures at the journal level}

The traditional measure of journal impact is the \textit{impact factor}.  Roughly, the impact factor of a journal is the  average number of recent citations received by articles published in the journal. More precisely, the impact factor of a journal for a specific census year is the mean number of citations that occurred in the census year to the articles published in the journal during a target window consisting of the two previous years. Such a measure was devised by Garfield, the founder of the Institute for Scientific Information (ISI). Today, Thomson-Reuters, that acquired the ISI in 1992, computes the the impact factor for journals it tracks and publishes it annually in the Journal Citation Reports (JCR) in separate editions for the sciences and the social sciences.

The impact factor has become a standard to evaluate the impact of journals. Nevertheless, the impact factor has many faults \cite{S97,AM00,C08}; the most commonly mentioned are:

\begin{itemize}

\item the target window (2 years) is too narrow for more theoretical disciplines, e.g., mathematics, in which results need to be well digested before they are cited;

\item the impact factor does not normalize for the differences in citation practices across different disciplines;

\item it does not represent a typical value of the number of citations to articles in the journal when the citation distribution is skewed (asymmetric), which is the usual case in bibliometrics (see Section \ref{distributions});

\item it counts citations without weighting them with the prestige of the citing journals.

\end{itemize}

It follows that impact factors highly vary across disciplines and over time \cite{AWBB09}. Moreover, due to the skewness of citation distributions and the fact that the impact factor is essentially a mean value, it is a (common) misuse of the impact factor to predict the importance of an individual publication, and hence of an individual researcher, based on the impact factor of the publication's journal. Indeed, most papers published in a high impact factor journal will ultimately be cited many fewer times than the impact factor may seem to suggest. Finally, some journals that have high impact factors are popular publication sources but are not appreciated by domain experts, that is, they are not prestigious sources \cite{BRS06}.

Nunes Amaral et al.\ propose a steady state version of the impact factor \cite{SSN08}. They
show that there exists a \textit{steady state period} of time specific to each journal such that the number of citations to paper published in the journal in that period will not significantly change in the future: poorly cited papers have stopped accruing citations, while the trickle of citations to highly cited ones is small when compared to the already accrued citations. Hence, there are journal-specific census and target windows that well characterize the final impact of journals; such windows highly diverge from the ones exploited in the impact factor computation.

Furthermore, the authors demonstrate that the \textit{logarithm} of the number of citations to papers published in a journal in its steady state period is approximately normally distributed and hence it has a well-defined typical value (the mean). The authors propose to use such a mean as an alternative impact metric to the commonly used 2-year impact factor. They show that the suggested ranking scheme strongly diverges from the 2-year impact factor one, but it is very similar to the \textit{probability ranking scheme}. The latter is the ranking that maximizes the probability that given a pair of papers $(a,b)$ from journals $A$ and $B$, respectively, paper $a$ is more cited than paper $b$ if $A$ is above $B$ in the ranking. The probability ranking is regarded as the \textit{optimal} ranking in the context of information retrieval \cite{JKR00}.

Both the impact factor and its steady version equally weights all citations: citations from highly reputed journals, like Nature, Science, and Proceedings of the National Academy of Sciences of USA, are treated as citations from more obscure ones. In other words, they are measures of popularity, but do not account for prestige. By contrast, the \textit{Eigenfactor}\texttrademark \ metric \cite{B07,BWW08,Eigenfactor} weights journal citations by the influence of the citing journals. As a result, a journal is influential if it is cited by other influential journals. The definition is clearly \textit{recursive} in terms of influence and the computation of the Eigenfactor scores involves the search of a \textit{stationary} distribution, which corresponds to the leading eigenvector of a perturbed citation matrix.

The Eigenfactor method was initially developed by Jevin West, Ben Althouse, Martin Rosvall, and Carl Bergstrom at the University of Washington and Ted Bergstrom at the University of California Santa Barbara. Eigenfactor scores are freely accessible at the Eigenfactor web site \cite{Eigenfactor} and, from 2007, they have been incorporated into Thomson-Reuters Journal Citation Reports (JCR) for both science and social science journals\footnote{Eigenfactor scores from 2007 are added to the Eigenfactor web site six months after they are published in JCR.}.

The idea underlying the Eigenfactor method originates from the work of \cite{PN76} in the field of bibliometrics  and from the contribution of \cite{H65} in the context of sociometry, which, in turn, generalizes Leontief's input-output model for the economic system \cite{L41}. Notably, Brin and Page use a similar intuition to design the popular PageRank algorithm that is part of their Google search engine: the importance of a web page is determined by the number of hyperlinks it receives from other pages as well as by the importance of the linking pages \cite{BP98,PBMW99}.

In the following, we illustrate the Eigenfactor method to measure journal influence as described at the Eigenfactor web site \cite{Eigenfactor}. The Eigenfactor computation uses a census citation window of one year and an earlier target publication window of five years. Let us fix a census year and let $C = (c_{i,j})$ be a journal-journal citation matrix such that $c_{i,j}$ is the number of citations from articles published in journal $i$ in the census year to articles published in journal $j$ during the target window consisting of the five previous years. Hence, the $i$th row represents the citations given by journal $i$ to other journals, and the $j$th column contains the citations received by journal $j$ from other journals. Journal self-citations are ignored, hence $c_{i,i} = 0$ for all $i$. Moreover, let $a$ be an article vector such that $a_i$ is the number of articles published by journal $i$ over the five-year target window divided by the total number of articles published by all journals over the same period. Notice that $a$ is normalized to sum to 1.

A dangling node is a journal $i$ that does not cite any other journals; hence, if $i$ is dangling, the $i$th row of the citation matrix has all 0 entries.  The citation matrix $C$ is transformed into a normalized matrix $H = (h_{i,j})$ such that all rows that are not dangling nodes are normalized by the row sum, that is, $$h_{i,j} = \frac{c_{i,j}}{\sum_{j} c_{i,j}}$$ for all non-dangling $i$ and all $j$. Furthermore, $H$ is mapped to a matrix $\hat{H}$ in which all rows corresponding to dangling nodes are replaced with the article vector $a$. Notice that $\hat{H}$ is row-stochastic, that is all rows are non-negative and sum to 1.

A new row-stochastic matrix $P$ is defined as follows: $$P = \alpha \hat{H} + (1- \alpha) A$$ where $A$ is the matrix with identical rows each equal to the article vector $a$, and $\alpha$ is a free parameter of the algorithm, usually set to $0.85$. Let $\pi$ be the left eigenvector of $P$ associated with the unity eigenvalue, that is, the vector $\pi$ such that $\pi = \pi P$. It is possible to prove that this vector exists and is unique. The vector $\pi$, called the \textit{influence vector}, contains the scores used to weight citations allocated in matrix $H$. Finally, the Eigenfactor vector $r$ is computed as $$r = 100 \cdot \frac{\pi H}{\sum_{i} [\pi H]_i}$$ That is, the Eigenfactor score of a journal is the sum of normalized citations received from other journals weighted by the Eigenfactor scores of the citing journals. The Eigenfactor scores are normalized such that they sum to 100.

The Eigenfactor metric has a solid mathematical background and an intuitive stochastic interpretation. The modified citation matrix $P$ is row-stochastic and can be interpreted as the transition matrix of a Markov chain on a finite set of states (journals). 
Hence, the influence vector $\pi$ corresponds to the stationary distribution of the associated Markov chain. Since $P$ is a primitive matrix, the Markov theorem applies, hence $\pi$ is the \textit{unique} stationary distribution and, moreover, the influence weight $\pi_j$ of the $j$th journal is the limit probability of being in state $j$ when the number of transition steps of the chain tends to infinity. Moreover, the Perron theorem for primitive matrices ensures that $\pi$ is a strictly positive vector corresponding to the leading eigenvector of $P$, that is, the eigenvector associated with the largest eigenvalue -- which is 1 because $P$ is stochastic. The described stochastic Markov process has an intuitive interpretation in terms of random walks on the citation network \cite{BWW08}. Imagine a researcher that moves from journal to journal by following chains of citations. The researcher selects a journal article at random and reads it. Then, he retrieves at random one of the citations in the article and proceeds to the cited journal. Hence, the researcher chooses at random an article from the reached journal and goes on like this. Eventually, the researcher gets bored of following citations, and selects a random journal in proportion to the number of article published by each journal. With this model of research, by virtue of the Ergodic theorem for Markov chains, the influence weight of a journal corresponds to the relative frequency with which the random researcher visits the journal.

The Eigenfactor score is a size-dependent measure of the total influence of a journal, rather than a measure of influence per article, like the impact factor. To make the Eigenfactor scores size-independent and comparable to impact factors, we need to divide the journal influence by the number of articles published in the journal. In fact, this measure, called Article Influence\texttrademark, is available both at the Eigenfactor web site and at Thomson-Reuters's JCR.

\section{Bibliometric data sources} \label{databases}

Bibliometric analysis can be conducted on the bases of any sufficiently large bibliographic database enhanced with citation counts. A bibliometric data source may be evaluated according to the following criteria:

\begin{itemize}
\item the coverage of the database;
\item the supported features for searching, sorting, and exporting bibliographic data as well as for computing performance indicators on them;
\item the availability of the database (free or subscription-based).
\end{itemize}

In particular, coverage is of crucial importance since it influences the outcomes of the computation for bibliometric indicators. An uneven coverage may produce performance measures that are too far from the real figures and this may lead to wrong decisions. Some aspects directly connected to database coverage are:

\begin{itemize}
\item what types (journals, conference papers, books, and so on) of works are covered and how evenly;
\item what research fields are covered and how evenly;
\item what languages other than English and what countries other than North American and Western European ones are covered and how evenly.
\end{itemize}

The bibliometric databases of the Institute for Scientific Information (ISI) have been the most generally accepted data sources for bibliometric analysis. The ISI was founded by Eugene Garfield in 1960. The ISI was acquired by Thomson in 1992, one of the world's largest information companies. In 2007, the Thomson Corporation reached an agreement with Reuters to combine the two companies under the name Thomson-Reuters (TR).

TR maintains Web of Knowledge, an online academic database which provides access to many resources, in particular:

\begin{itemize}
\item Web of Science (WoS), which includes the Science Citation Index (SCI), the
Social Science Citation Index (SSCI), and the Arts and Humanities Citation Index (AHCI);

\item
Journal Citation Reports (JCR), containing citation information, and in particular the impact factor, for the journals tracked by TR. JCR are published annually in separate editions for the sciences and the social sciences.

\end{itemize}

The use as TR citation databases, in particular as the only bibliographic source for bibliometric analysis, attracted quite a number of critics. The most mentioned flaws are:

\begin{enumerate}

\item it provides different coverage between research fields;

\item it is limited to citations from journals but does not count citations from other sources, mainly from books and most conference proceedings;

\item it covers mainly North American, Western European, and English-language titles;

\item is only available to those academics whose institutions are able and willing to bear the subscription cost.
\end{enumerate}

Flaw number one is particulary serious since it is the main cause of the variation of the impact factor measure across fields \cite{AWBB09}. The \textit{internal coverage} of a bibliometric data source  with respect to a field is defined as the fraction of citations coming from papers internal to the database and belonging to the field that match a paper in the same data source. The internal coverage highly varies across disciplines, e.g.\ 0.803 for molecular and cell biology, 0.552 for mathematics, and 0.226 for computer science.  This means that, for instance, more than 3/4 of the citations from computer science papers indexed in WoS are addressed to papers that are \textit{not} contained in WoS. Furthermore, drawback number 2 is critical for disciplines like computer science that heavily rely on conference publications and for humanities whose scholars frequently publish books.

Two major alternatives to Web of Science are Elsevier's Scopus and Google Scholar. Scopus, as Web of Science, is a subscription-based proprietary databases. On the contrary, Google Scholar is freely accessible. There are many studies that compare citation data retrieved on different data sources. Table~\ref{relwork} displays some of these. The first column shows the publication reference of the study, the second column contains the set of data sources that are compared, while the research field of the publications considered in the study is given in the third column. The papers are sorted in chronological order.

\begin{table}
\begin{center}
\begin{tabular}{|l|p{5cm}|p{5cm}|}
\hline
\textbf{ref} &	\textbf{compared sources} &	\textbf{field} \\ \hline
\cite{GMLG01} & CiteSeer and Web of Science & computer science  \\ \hline
\cite{ZL02} & CiteSeer and  Web of Science & computer science  \\ \hline
\cite{W02} & Chemical Abstracts and Web of Science & chemistry \\ \hline
\cite{BB05} & Web of Science, Scopus, and Google Scholar   & JASIST papers  \\ \hline
\cite{J05} &  Web of Science, Scopus, and Google Scholar & Current Science papers and Eugene Garfield papers
\\ \hline
\cite{N05} & Web of Science and Google Scholar & webometrics  \\ \hline
\cite{PS05} & Web of Science and Google Scholar & different disciplines \\ \hline
\cite{BBGW06} & Web of Science, Scopus, and Google Scholar & oncology and
condensed matter physics  \\ \hline
\cite{KT07} & Web of Science and Google Scholar & different disciplines  \\ \hline
\cite{S06} & Web of Science and Google Scholar  & business sciences  \\ \hline
\cite{NO07} & CSA Illumina, Web of Science, Scopus, and Google Scholar & social sciences \\ \hline
\cite{BLL07} & Web of Science, Scopus, and Google Scholar & different disciplines   \\ \hline
\cite{MY07} & Web of Science, Scopus, and Google Scholar  & library and information science
\\ \hline
\cite{KT08} & Web of Science and Google Scholar & different disciplines  \\ \hline
\cite{SV08} & Web of Science and Google Scholar  & information science  \\ \hline
\cite{MR08} & Web of Science, Scopus and Google Scholar & human-computer interaction  \\ \hline
\cite{B08} & Web of Science, Scopus, and Google Scholar & different disciplines  \\ \hline
\cite{S08} & Web of Science, Scopus and Google Scholar & library and information science and information retrieval  \\ \hline
\cite{BMSRTD09} & Web of Science, Scopus, Google Scholar, and Chemical Abstracts & chemistry  \\ \hline

\end{tabular}
\end{center}
\caption{Literature comparing citation data over different data sources.}
\label{relwork}
\end{table}

The rest of this section illustrates a large-scale comparison between Web of Science, Scopus and Google Scholar conducted by Meho and Yang in 2007 \cite{MY07}. The study covers more than 10000 citations to approximately 1100 scholarly works of all 15 faculty members of the School of Library and Information Science (LIS) at Indiana University-Bloomington. The authors found  that both Web of Science and Scopus provide substantial factual information about the database, including the number of records and the list of titles indexed. On the contrary, Google Scholar refuses to publish information about its coverage and frequency of updates.

Moreover, both Web of Science and Scopus offer features for searching, sorting, and exporting the bibliographic data. On the contrary, Google Scholar does not provide the retrieved data in a useful bibliographic format and does not allow sorting it in any way. Collecting (extracting, verifying, cleaning, organizing, classifying, and saving into a bibliographic format) data from Google Scholar took the authors 30 as much time as collecting Web of Science data and 15 as much time as as collecting Scopus data.

The authors analyzed the citations to works of LIS members published between 1996 and 2005 divided by document type. For Web of Science and Scopus, most of these citations come from journals (88.7\% and 84.4\%, respectively), and a few from conference papers (11.3\% and 15.6\%, respectively). On the contrary, Google Scholar indexes more types of works, including journal papers (42.5\%), conference papers (33.7\%), theses (9.8\%), books (5.5\%), reports (4.8\%) and other document types (3.7\%);

Furthermore, the authors studied the distribution of unique and overlapping citations to works of LIS members published between 1996 and 2005 that were found in \textit{journal and conference articles only}. Web of Science contains 2023 citations, Scopus contains 2301 citations, and Google Scholar contains 4181 citations. The overlap of citations between the three databases is relatively low, with significant differences from one research area to another: the overlapping between Web of Science and Scopus is 1591 citations (58.2\%) out of the 2733 citations found in both the databases; Web of Science misses 710 citations (26\%) of those of Scopus while Scopus misses 432 citations (15.8\%) of those of Web of Science.

The overlapping between Google Scholar and the union of Web of Science and Scopus is 1629 citations (30.8\%) out of the 5285 citations found in all the three databases; Google Scholar misses 1104 citations (20.9\%) of those found in the union of Web of Science and Scopus, and the union of Web of Science and Scopus misses 2552 citations (48.3\%) of those found in Google Scholar.

The former figure is quite striking, since virtually all citations of Web of Science and Scopus come from referred reputable sources. Rumors are that some publishers did not allow Google Scholar crawlers to enter their databases (notably, Elsevier and American Chemical Society).
As for the second figure, the authors noticed that:

\begin{itemize}
\item most of the citations uniquely found by Google Scholar are from refereed sources;

\item most of these citations come from low impact sources;

\item most of these citations were identified trough documents made available online by their authors rather than from the source's official site.
\end{itemize}

The authors studied the distribution of citations by language and found that Google Scholar provides better coverage of non-English language materials (6.9\%) with respect to both Web of Science (1.1\%) and Scopus (0.7\%).

Meho and Yang concluded that Web of Science, Scopus, and Google Scholar complement rather than replace each other, so they should be used together rather than separately in citation analysis. In particular, although Web of Science remains an indispensable citation database, it should not be used alone for locating citations, because both Scopus and Google Scholar identify a considerable number of citations not found in Web of Science. Although Google Scholar unique citations are not of the same quality of those found in the two proprietary databases, they could be useful in showing evidence of broader international impact.

The authors also concluded that there is an important impact advantage in favor of the articles, and the corresponding journals, that their authors make available online (on personal web pages or on electronic preprints archives like arXiv) since they are more likely discovered by human and automatic agents (like crawlers of Google Scholar), possibly increasing the citation impact.

\section{Bibliometric distributions} \label{distributions}

The probability distributions that are usual suspects in bibliometrics are Pareto, (stretched) exponential, and lognormal distributions.

\medskip
\noindent
\textit{Pareto distribution}. Also known as power law distribution, it has been used to model phenomena where most of the effects come from few of the causes. The distribution is named after the Italian economist Vilfredo Pareto who originally observed it studying the allocation of wealth among individuals: a larger share of wealth of any society (approximately 80\%) is owned by a smaller fraction (about 20\%) of the people in the society~\cite{P97}. Examples of phenomena that are approximately Pareto distributed are: size of human settlements, size of meteorites, standardized price returns on individual stocks, file size of Internet traffic using TPC protocol, duration of transactions in database management systems, word frequency in relatively lengthy texts (Zipf law~\cite{Z49}), and scientific productivity of scholars (Lotka law~\cite{L26}). Furthermore, in 1998 Redner analyzed the citation distribution for all papers in journals which were catalogued by the ISI at that time; the author found that the asymptotic tail of the citation distribution appears to be described by a Pareto law~\cite{R98}.

The probability density function for a Pareto distribution is defined for $x \geq 1$ in terms of parameter $\alpha > 0$ as follows:
$$
f(x) = \frac{\alpha}{x^{\alpha + 1}}
$$
The cumulative distribution function is:
$$
F(x) = 1 - \frac{1}{x^{\alpha}}
$$
The mean is $\alpha / (\alpha -1)$ for $\alpha > 1$, and infinite otherwise. The median is $\sqrt[\alpha]{2}$ and the mode is $1$. Notice that the mean is greater than the median which is greater than the mode and the limit for $\alpha \rightarrow \infty$ of both the mean and the median is the mode $1$. Skewness is $$\gamma = \frac{2 (1 + \alpha)}{\alpha - 3} \sqrt{\frac{\alpha - 2}{\alpha}}$$ for $\alpha > 3$, and kurtosis  is $$\kappa = \frac{6(\alpha^3 + \alpha^3 - 6 \alpha - 2)}{\alpha (\alpha-3) (\alpha-4)}$$ for $\alpha > 4$. Both skewness and kurtosis are greater than zero and tend to 2 and 6, respectively, as $\alpha \rightarrow \infty$. The raw moments are found to be $E(X^n) = \alpha  / (\alpha - n)$ for $\alpha > n$.

\medskip
\noindent
\textit{Stretched exponential distribution}. This is a family of extensions of the well-known exponential distribution characterized by fatter tails. Laherr\'{e}re and Sornette showed that different phenomena in nature and economy can be described in the regime of the exponential distribution, including radio and light emission from galaxies, oilfield reserve size, agglomeration size, stock market price variation, biological extinction event, earthquake size, temperature variation of the earth, and, notably, citation of the most cited physicists in the world~\cite{LS98}.

The probability density function is a simple extension of the exponential distribution with one additional stretching parameter $\alpha$:
$$
f(x) = \alpha \lambda^\alpha x^{\alpha -1} e^{-(\lambda x)^\alpha}
$$
where $x \geq 0$, $\lambda > 0$ and $0 < \alpha \leq 1$. In particular, if the stretching parameter $\alpha = 1$, then the distribution is the usual exponential distribution. When the parameter $\alpha$ is not bounded from $1$, the resulting distribution is better known as the Weibull distribution. The cumulative distribution function is:
$$
F(x) = 1 - e^{-(\lambda x)^\alpha}
$$
It can be shown that the $n$th raw moment $E(X^n)$ is $\frac{1}{\lambda^n} \Gamma(\frac{\alpha + n}{\alpha})$, where $\Gamma(x)$ is the Gamma function, an extension of the factorial function to real and complex numbers, defined by:
$$\Gamma(x) = \int_{0}^{\infty} t^{x-1} e^{-t} dt$$
In particular, it holds that $\Gamma(1) = 1$ and $\Gamma(x+1) = x \Gamma(x)$. Hence, for a positive integer $n$, we have $\Gamma(n+1) = n!$. Notice that, if $\alpha = 1$, then the raw moments are given by $n! / \lambda^n$ and they correspond to the raw moments of the exponential distribution. In particular, the mean $E(X)$ is the first raw moment that is equal to $\frac{1}{\lambda} \Gamma(\frac{\alpha + 1}{\alpha})$. The median is given by $\frac{1}{\lambda} \sqrt[\alpha]{\log 2}$ and the mode is 0. Notice again that the mean is greater than the median that is greater than the mode.

\medskip
\noindent
\textit{Lognormal distribution}. It is the distribution of any random variable whose logarithm is normally distributed. A lognormal distribution characterizes phenomena determined by the multiplicative product of many independent effects. The lognormal distribution is a usual suspect in bibliometrics. In a 1957 study based on the publication record of the scientific research staff at Brookhaven National Laboratory, Shockley observed that the scientific publication rate is approximately lognormally distributed~\cite{S57}. More recently, Stringer et al.\ studied the citation distribution for journals indexed in Web of Science publishing at least 50 articles per year for at least 15 years and demonstrated that in a steady citational state the logarithm of the number of citations has a journal-specific typical value~\cite{SSN08}. Finally, Radicchi et al.\ analyzed the distribution of the ratio between the number of citations received by an article and the average number of citations received by articles published in the same field and year for papers in different research categories (the category closest to computer science is cybernetics)~\cite{RFC08}. They found a similar distribution for each category with a good fit with the lognormal distribution.

The lognormal probability density function is defined in terms of parameters $\mu$ and $\sigma > 0$ as follows:
$$
f(x) = \frac{1}{x \sigma \sqrt{2 \pi}} e^{-\frac{(\log(x)-\mu)^2}{2 \sigma^2}}
$$
for $x > 0$. The parameters $\mu$ and $\sigma$ are the mean and standard deviation of the variable's natural logarithm. The cumulative distribution function has no closed-form expression and is defined in terms of the density function as for the normal distribution.  The mean is $e^{\mu + \sigma^2 / 2}$, the median is $e^{\mu}$ and the mode is $e^{\mu - \sigma^2}$. Notice that the mean is greater than the median which is greater than the mode. This suggests a positive asymmetry of the distribution. Indeed, skewness is $(e^{\sigma^2} - 1) \sqrt{e^{\sigma^2} - 1} > 0$. Moreover, excess kurtosis is $e^{4 \sigma^2} + 2 e^{3 \sigma^2} + 3 e^{2 \sigma^2} - 6 > 0$. The raw moments are given by $E(X^n) = e^{n \mu + n^2 \sigma^2 / 2}$.

It is interesting to observe that all the above distributions that are commonly used to model  bibliometric phenomena are positively (right) skewed. A distribution is symmetric if the values are equally distributed around a typical figure (the mean);  a well-known example is the normal (Gaussian) distribution. A distribution is right-skewed if it contains many low values and a relatively few high values. It is left-skewed if it comprises  many high values and a relatively few low values. As a rule of thumb, when the mean is larger than the median the distribution is right-skewed and when the median dominates the mean the distribution is left-skewed. A more precise numerical indicator of distribution skewness is the third standardized central moment, that is $$\gamma = \frac{E[(X - \mu)^3]}{\sigma^3}$$
where $\mu$ and $\sigma$ are mean and standard deviation of distribution of random variable $X$, respectively. A value close to 0 indicates symmetry; a value greater than 0 corresponds to right skewness, and a value lower than 0 means left skewness.

The observed right skewness might be considered as an application of the more general Pareto Principle (also known as 80-20 rule) [7]. The principle states that:

\begin{quote}
\textit{Most (approximately 80\%) of the effects comes from few (about 20\%) of the causes.}
\end{quote}

It has been suggested that the irreducible skewness of distributions of scholar productivity and article citedness may be explained by sociological reinforcement mechanisms such as the Principle of Cumulative Advantage. De Solla Price formulated this in 1976 as follows \cite{P76}:

\begin{quote}
\textit{Success seems to breed success. A paper which has been cited many times is more likely to be cited again than one which has been little cited. An author of many papers is more likely to publish again than one who has been less prolific. A journal which has been frequently consulted for some purpose is more likely to be turned to again than one of previously infrequent use.}
\end{quote}

The Matthew Effect may additionally contribute to skewness. According to Merton \cite{M68}:

\begin{quote}
\textit{The Mathew Effect consists in the accruing of greater increments of recognition for particular scientific contributions to scientists of considerable repute and the withholding of such recognition from scientists who have not yet made their mark.}
\end{quote}

It takes the name from the following line in Jesus' parable of the talents in the biblical Gospel of Mathew: \textit{For unto every one that hath shall be given, and he shall have abundance: but from him that hath not shall be taken away even that which he hath.}

Interestingly, Seglen claims that skewness is an intrinsic characteristic of distributions related to extreme types of human efforts; although scientific ability may be normally distributed in the general population, scientists are likely to form an extreme-property distribution to their speciality be in terms of citedness or in terms of productivity. This statistical pattern is expected for different types of highly specialized human activity, a parallel being found in the distribution of performance by top athletes~\cite{S92}.

\section{Bibliometric maps} \label{maps}

A first and crucial step is the building of a map is definition of a research field. There are two main approaches: concept-similarity mapping and citation mapping. The concept-similarity approach defines a research field on the basis of repeated concepts (keywords) in publications \cite{CCTB83,C94}. It can be further divided in the two following methods: co-publication analysis, in which publications are related if they mention the same concepts, and  co-concept analysis, in which concepts are related if they are mentioned together in the same publication.

The citation approach clusters a research field on the basis of citations in publications. Two typical methods to identify similar publications are co-citation coupling \cite{S73}, in which publications are related when they are cited by the same papers, and bibliographic coupling \cite{K63}, in which papers are related when they cite the same papers. In the following, we introduce these two techniques with the help of a model suggested in \cite{SG83} and refined in \cite{GC96}. Suppose we have $n$ publications $p_1, \ldots, p_n$ that cite $m$ references $r_i, \ldots, r_m$. We build a Boolean citation matrix $C = (c_{i,j})$ of size $n \times m$ such that $c_{i,j} = 1$ if $p_i$ cites $r_j$ and $c_{i,j} = 0$ otherwise.  Let $c_i = \sum_j c_{i,j}$ be the number of cited references of $p_i$ and $c^j = \sum_i c_{i,j}$ be the number of citations received by $r_j$. A measure of bibliographic coupling between publications $p_i$ and $p_j$ is:
$$r_{i,j} = \frac{\sum_k c_{i,k} \cdot c_{j,k}}{\sqrt{c_i \cdot c_j}}$$
This is the ratio of the number of references shared by publications $p_i$ and $p_j$ and the geometric mean of the number of references of the two papers concerned. Notice that $0 \leq r_{i,j} \leq 1$, and $r_{i,j} = 0$ when publications $p_i$ and $p_j$ share no references, while $r_{i,j} = 1$ when publications $p_i$ and $p_j$ have the same bibliography. Geometrically, $r_{i,j}$ is the cosine of the angle formed by the $i$th and $j$th rows of the citation matrix, which is $0$ when the two vectors are orthogonal, and is $1$ when they are parallel. In matrix notation, let $A = (a_{i,j}) = C C^T$, that is, $a_{i,j}$ is the number of references shared by $i$th and $j$th publications, and, in particular, $a_{i,i} = c_i$ is the number of references of $p_i$. Let $D$ be the diagonal matrix such that the $i$th diagonal entry is $1 / \sqrt{a_{i,i}}$. Then we have that
$R = (r_{i,j})$ is defined as $$R = D A D$$

On the other hand, a measure of co-citation coupling between publications $p_i$ and $p_j$ is:
$$s_{i,j} = \frac{\sum_k c_{k,i} \cdot c_{k,j}}{\sqrt{c^i \cdot c^j}}$$
This is the ratio of the number of articles that cite both publications $p_i$ and $p_j$ and the geometric mean of the number of citations received by the two publications involved. This is also the cosine of the angle formed by the $i$th and $j$th columns of the citation matrix.
Again, $0 \leq s_{i,j} \leq 1$, and $s_{i,j} = 0$ when publications $p_i$ and $p_j$ are never co-cited, while $s_{i,j} = 1$ when publications $p_i$ and $p_j$ are always cited together. In matrix notation, let $B = (b_{i,j}) = C^T C$, that is, $b_{i,j}$ is the number of articles that co-cited  $i$th and $j$th publications, and, in particular, $b_{i,i} = c^i$ is the number of citations gathered by $p_i$. Let $D'$ be the diagonal matrix such that the $i$th diagonal entry is $1 / \sqrt{b_{i,i}}$. Then we have that $S = (s_{i,j})$ is defined as $$S = D' B D'$$

It is worth noticing that the similarity formulas used in citation coupling closely resemble Pearson correlation coefficient formula for two statistical samples $x$ and $y$, that is:
$$r_{xy} = \frac{\sigma_{xy}}{\sigma_x \cdot \sigma_y} =
\frac{\sum_{k} (x_{k} - \mu_x) \cdot (y_{k} - \mu_y)}{\sqrt{\sum_{k} (x_{k} - \mu_x)^2} \cdot \sqrt{\sum_{k} (y_{k} - \mu_y)^2}}$$
In particular, when the means of both statistical samples $x$ and $y$ are null, the Pearson correlation coefficient is exactly the cosine of the angle formed by the two sample vectors and the two measures coincide.

Once the similarity strength between bibliometric units has been established, bibliometric units are typically represented as graph nodes and the similarity relationship between two units is represented as a weighted edge connecting the units, where weights stand for the similarity intensity. Such visualizations are called \textit{bibliometric maps}. Such maps are powerful but they are often highly complex. It therefore is helpful to abstract the network into inter-connected modules of nodes. Good abstractions both simplify and highlight the underlying structure and the relationships that they depict. When the units are publications or concepts, the identified modules represent in most cases recognizable research fields. In the rest of this section, we describe three methods for creating these abstractions: clustering, principal component analysis, and information-theoretic abstractions.

\subsection{Clustering}

Informally, clustering is the process of organizing objects into groups whose members are similar in some way~\cite{A73,JD88}. A cluster is a collection of objects which are similar between them and are dissimilar to objects belonging to other clusters. Clustering can be formalized as follows. We are given a weighted undirected graph $G$, where the weight function assigns a dissimilarity value to pair of nodes, and an objective function $f$ that assigns a value of merit to any partition of the set of nodes of $G$. Clustering problems are optimization problems that usually have one of the following forms~\cite{G85}:

\begin{itemize}

\item Let $G$ be a graph, $f$ be an objective function, and $k$ be an integer. Find a partition of nodes in $G$ with cardinality $k$ and with the least value for the objective function.

\item Let $G$ be a graph, $f$ be an objective function, and $c$ be a real number. Find the smallest partition of nodes in $G$ with objective function value less than or equal to the value $c$.

\end{itemize}

The first type of clustering problem is usually approached using \emph{repeated partition techniques}. These techniques choose an initial partition with $k$ clusters and then move objects between clusters trying to minimize the objective function. The procedure stops as soon as a \emph{local} minimum for the objective function is reached. The most popular algorithm in this category is K-means~\cite{M67}.

\emph{Hierarchical clustering methods} are typically applied to solve the second type of clustering problem~\cite{J67}. In this case, the size of the partition is not fixed in advance. These algorithms are of two kinds: agglomerative and divisive. An agglomerative strategy starts with a singleton partition containing a cluster for each object and then merges similar clusters until the universal partition is obtained. A divisive strategy starts from the universal partition containing a unique set with all objects and then divides clusters that include dissimilar objects until the singleton partition is reached. Both methods can use different methods to decide what clusters to join or to divide. They output a hierarchical structure (a dendrogram) describing the whole merging/dividing process. This structure can be used to choose the smallest partition among the generated ones (a small subset of all partitions) with objective function value less than or equal to the given threshold.

The computational complexity of clustering problems mainly depends on the properties of the weight function that measures the distance between two objects and on the objective function that evaluates the goodness of a given partition of the space. Many exact and approximated clustering problems are known to be hard to solve, in particular NP-hard~\cite{G85,GS76}. Hence a polynomial strategy cannot guarantee to find the optimum solution.

\subsection{Principal component analysis}

Principal component analysis (PCA)~\cite{P01,J02} is a multivariate statistic method used to reduce a multi-dimensional  space to a lower dimension.

Given a set of correlated variables $X = \{X_1, X_2, \ldots ,X_n\}$, the aim of PCA is to find new artificial variables $Y = \{Y_1, Y_2, \ldots ,Y_m\}$, with $m < n$, such that (i) each new variable $Y_i$ is obtained as a linear combination of the original variables  (ii) the new variables $Y_i$ are pairwise uncorrelated, (iii) the variance of $Y_i$ decreases as the index $i$ increases, and (iv) the sum of the variance of the new variables $Y_i$ is a significant portion of the sum of the variance of the original variables $X_i$. The principal components $Y_i$ represent the most informative orthogonal aspects of the data set.

A simple method to find principal components is the following:

\begin{enumerate}

\item find the covariance matrix $\Sigma_X$ of variables in $X$;

\item compute the eigenvalues $\lambda_i$ of $\Sigma_X$ and sort them in descending order: $\lambda_1 \geq \lambda_2 \geq \ldots \lambda_n \geq 0$;

\item find the eigenvectors $e_i$ associated with the eigenvalues $\lambda_i$. The $i$-th principal component $Y_i$ is $\sum_{j=1}^{n} e_{i,j} X_j$, where $e_{i,j}$ is the $j$-th component of vector $e_{i}$;

\item determine the $m \leq n$ most informative principal components $Y_1, Y_2, \ldots ,Y_m$ such that: $$\frac{\lambda_1 + \lambda_2 + \ldots + \lambda_m}{\lambda_1 + \lambda_2 + \ldots + \lambda_n} \geq \alpha$$ where $0 < \alpha \leq 1$ is a threshold (often fixed at $0.8$).

\end{enumerate}

An alternative method (Kaiser method) to isolate the $m$ principal components is to choose those components such that $\lambda_i > 1$. Notice that, since $\Sigma_X$ is semi-definite positive, its eigenvalues are greater than or equal to 0. Moreover, it holds that the variance $var(Y_i) = \lambda_i$ and $\sum_{i=1}^{n} var(X_i) = tr(\Sigma_X) = \sum_{i=1}^{n} \lambda_i =  \sum_{i=1}^{n} var(Y_i)$. Hence, the most informative principal components contribute at least a fraction of $\alpha$ to the total variance of the original data set $X$.
If variables in $X$ have different units of measure, then they must be standardized before applying the method. This is equivalent to work with the correlation matrix $R_X$ instead of with the covariance matrix $\Sigma_X$.

The contribution of the original variable $X_j$ to the new variable $Y_i$ is given by the eigenvector component $e_{i,j}$. The highest is $e_{i,j}$ in absolute value, the highest is the contribution of $X_j$ to $Y_i$. Moreover, it holds that the correlation between $Y_i$ and $X_j$ is $e_{i,j} \sqrt{\lambda_i} / sd(X_j)$, where $sd(X_j)$ is the standard deviation of $X_j$. Hence the sign of $e_{i,j}$ gives the sign of the correlation between $Y_i$ and $X_j$. It turns out that variables $X_j$ can be clustered by associating each $X_j$ to the component $Y_i$ such that the two variables are most correlated.

\subsection{Information-theoretic abstractions}

Rosvall and Bergstrom \cite{RB07} propose a model for resolving community structure in complex networks based on information theory. They start from the following observation: when we describe a network as a set of interconnected modules, we are highlighting certain regularities of the network's structure while filtering out the relatively unimportant details. Thus, a modular description of a network can be viewed as a lossy compression of that network's topology. The best maps are those that convey a great deal of information while requiring minimal bandwidth; i.e., they are good compressions. This view suggests that we can approach the challenge of identifying the community structure of a complex network as a  problem in information theory. The authors envision the process of abstraction of a complex network as a communication process. The link structure of the network is a random variable $X$; this is compressed into a simplified description $Y$ which is sent through a noiseless communication channel. The receiver uses the abstraction $Y$ to make guesses $Z$ about the structure of the original network $X$. The partition of the original network is achieved by minimizing the length $L(Y)$ of the abstraction $Y$ plus the length $L(X|Y)$ of the additional information that is necessary to describe the original network $X$ given its simplified representation $Y$. The minimization problem is tackled using the  simulated annealing approach.

In a successive work \cite{RB08}, the same authors use ergodic random walks on complex directed weighted networks to reveal community structure. The intuition here is as following. The local interactions among the subunits of a network system induce a system-wide flow of information that characterizes the behavior of the whole system. Consequently, if we want to understand how network structure relates to system behavior, we need to understand the flow of information on the network. A group of nodes among which information flows quickly and easily can be aggregated and described as a single well connected module; the links between modules capture the avenues of information flow between those modules. The authors use an infinite random walk as a proxy of the information flow and identify the modules that compose the network by minimizing the expected description length of the ergodic random walk within and across the modules. This is the sum of the entropy of the movements across modules and of the entropy of movements within modules. Huffman code \cite{H52} is exploited to encode the random walk by assigning short codewords to frequently visited nodes and modules, and longer codewords to rare ones, much as common words are short in spoken language \cite{Z49}. Shannon source coding theorem \cite{S48} provides a lower bound to the average length of a codeword. The minimization problem is approached using a greedy search algorithm and the solution is refined with the aid of simulated annealing.

\section{Quotations} \label{quote}

\begin{quote}
\emph{``Measuring is knowing'' --  Heike Kamerlingh Onnes}
\end{quote}

\begin{quote}
\emph{``Not everything that can be counted counts, and not everything that counts can be counted'' -- Albert Einstein}
\end{quote}

\begin{quote}
\emph{``If scientometrics is a mirror of science in action, then scientometricians' particular responsibility is to both polish the mirror and warn against optical illusions'' -- Michel Zitt}
\end{quote}

\begin{quote}
\emph{``No amount of fancy statistical footwork will overcome basic inadequacies in either the appropriateness or the integrity of the data collected'' -- Goldstein and Spiegelhalter}
\end{quote}

\begin{quote}
\emph{``We think of statistics as facts that we discover, not numbers we create'' -- Joel Best}
\end{quote}

\begin{quote}
\emph{``Citations are frozen footprints in the landscape of scholarly achievement which bear witness of the passage of ideas'' -- Blaise Cronin}
\end{quote}

\begin{quote}
\emph{``The use of a single index crashes the multidimensional space of bibliometrics into one single dimension'' -- Wolfgang Gl\"{a}nzel}
\end{quote}

\begin{quote}
\emph{``For unto every one that hath shall be given, and he shall have abundance: but from him that hath not shall be taken away even that which he hath'' -- Jesus of Nazareth}
\end{quote}


\begin{thebibliography}{10}

\bibitem{C73}
A.~de~Candolle.
\newblock {\em Histoire des sciences et des savants depuis deux si\`{e}cles}.
\newblock Genève/Basel: H. Georg, 1873.

\bibitem{S63}
D.~de~Solla~Price.
\newblock {\em Little Science, Big Science}.
\newblock Columbia University Press, New York, 1963.

\bibitem{P69}
A.~Pritchard.
\newblock Statistical bibliography or bibliometrics?
\newblock {\em Journal of Documentation}, 24:348--349, 1969.

\bibitem{L26}
A.~J. Lotka.
\newblock The frequency distribution of scientific productivity.
\newblock {\em Journal of the Washington Academy of Sciences}, 16:317--323,
  1926.

\bibitem{B34}
S.~C. Bradford.
\newblock Sources of information on specific subjects.
\newblock {\em Engineering}, 137:85--86, 1934.

\bibitem{Z49}
G.~K. Zipf.
\newblock {\em Human behavior and the principle of least effort: An
  introduction to human ecology}.
\newblock Addison-Wesley, Cambridge, MA, 1949.

\bibitem{vR06b}
A.~F.~J. van Raan.
\newblock Measuring science. {C}apita selecta of current main issues.
\newblock In H.~F. Moed, W.~Gl\"{a}nzel, and U.~Schmoch, editors, {\em Handbook
  of quantitative science and technology research. The use of publication and
  patent statistics in studies of S\&T systems}, pages 19--50. Kluwer Academic
  Publishers, 2006.

\bibitem{S81}
L.~C. Smith.
\newblock Citation analysis.
\newblock {\em Library Trends}, 30(1):85, 1981.

\bibitem{C81}
B.~Cronin.
\newblock The need for a theory of citation.
\newblock {\em Journal of Documentation}, 37:16--24, 1981.

\bibitem{MM89}
M.~H. MacRoberts and B.~R. MacRoberts.
\newblock Problems of citation analysis: A critical review.
\newblock {\em Journal of the American Society for Information Science},
  40(5):342--349, 1989.

\bibitem{S92}
P.~O. Seglen.
\newblock The skewness of science.
\newblock {\em Journal of the American Society for Information Science},
  43(9):628--638, 1992.

\bibitem{W88}
L.~C.~H. Westney.
\newblock Historical rankings of science and technology: A citationist
  perspective.
\newblock {\em The Journal of the Association for History and Computing}, 1(1),
  1988.

\bibitem{K99}
J.~M. Kleinberg.
\newblock Authoritative sources in a hyperlinked environment.
\newblock {\em Journal of the ACM}, 46(5):604--632, 1999.

\bibitem{PBMW99}
Brin, L.~Page, R.~Motwani, and T.~Winograd.
\newblock The {PageRank} citation ranking: Bringing order to the {Web}.
\newblock Technical Report 1999-66, Stanford InfoLab, November 1999.
\newblock Retrieved July 1, 2009, from
  \texttt{http://ilpubs.stanford.edu:8090/422/}.

\bibitem{N94}
F.~Narin.
\newblock Patent bibliometrics.
\newblock {\em Scientometrics}, 30(1):147--155, 1994.

\bibitem{LLS98}
W.~M. Landes, L.~Lessig, and M.~E. Solimine.
\newblock Judicial influence: A citation analysis of federal courts of appeal
  judges.
\newblock {\em Journal of Legal Studies}, 27:271--332, 1998.

\bibitem{MT05}
H.~Murai and A.~Tokosumi.
\newblock A network analysis of hermeneutic documents based on {Bible}
  citations.
\newblock In {\em Annual Conference of the Cognitve Science Society}, pages
  1565--1570, 2005.

\bibitem{vR06c}
A.~F.~J. van Raan.
\newblock Comparison of the {H}irsch-index with standard bibliometric
  indicators and with peer judgment for 147 chemistry research groups.
\newblock {\em Scientometrics}, 67(3):491--502, 2006.

\bibitem{G06}
W.~Gl\"{a}nzel.
\newblock On the opportunities and limitations of the h-index.
\newblock {\em Science Focus}, 1(1):10--11, 2006.

\bibitem{AM00}
M.~Amin and M.~Mabe.
\newblock Impact factors: use and abuse.
\newblock {\em Perspectives in publishing}, 1:1--6, 2000.

\bibitem{RFC08}
F.~Radicchi, S.~Fortunato, and C.~Castellano.
\newblock Universality of citation distributions: Toward an objective measure
  of scientific impact.
\newblock {\em Proceedings of the National Academy of Sciences of USA},
  105(45):17268--17272, 2008.

\bibitem{H05}
J.~E. Hirsch.
\newblock An index to quantify an individual's scientific research output.
\newblock {\em Proceedings of the National Academy of Sciences of USA},
  102(46):16569--16572, 2005.

\bibitem{B05}
P.~Ball.
\newblock Index aims to fair ranking of scientists.
\newblock {\em Nature}, 436:900, 2005.

\bibitem{Ba07}
P.~Ball.
\newblock Achievement index climbs the ranks.
\newblock {\em Nature}, 448:737, 2007.

\bibitem{H07}
J.~E. Hirsch.
\newblock Does the h index have predictive power?
\newblock {\em Proceedings of the National Academy of Sciences of USA},
  104(49):19193--19198, 2007.

\bibitem{BD07}
L.~Bornmann and H-D. Daniel.
\newblock What do we know about the h index?
\newblock {\em Journal of the American Society for Information Science and
  Technology}, 58(9):1381--1385, 2007.

\bibitem{BCK06}
P.~D. Batista, M.~G. Campiteli, and O.~Konouchi.
\newblock Is it possible to compare researchers with different scientific
  interests?
\newblock {\em Scientometrics}, 68(1):179--189, 2006.

\bibitem{E06}
L.~Egghe.
\newblock Theory and practice of the g-index.
\newblock {\em Scientometrics}, 69(1):131--152, 2006.

\bibitem{J06}
B.~H. Jin.
\newblock {H-index}: An evaluation indicator proposed by scientist.
\newblock {\em Science Focus}, 1(1):8--9, 2006.

\bibitem{KMS07}
C.~Katsaros, Y.~Manolopoulos, and A.~Sidiropoulos.
\newblock Generalized h-index for disclosing latent facts in citation networks.
\newblock {\em Scientometrics}, 72(2):253--280, 2007.

\bibitem{S97}
P.~O. Seglen.
\newblock Why the impact factor of journals should not be used for evaluating
  research.
\newblock {\em British Medical Journal}, 314:498--502, 1997.

\bibitem{C08}
P.~Campbell.
\newblock Escape from the impact factor.
\newblock {\em Ethics in science and environmental politics}, 8:5--7, 2008.

\bibitem{AWBB09}
B~M. Althouse, J.~D. West, C.~T. Bergstrom, and T.~Bergstrom.
\newblock Differences in impact factor across fields and over time.
\newblock {\em Journal of the American Society for Information Science and
  Technology}, 60(1):27--34, 2009.

\bibitem{BRS06}
J.~Bollen, M.~A. Rodriguez, and H.~Van de~Sompel.
\newblock Journal status.
\newblock {\em Scientometrics}, 69(3):669--687, 2006.

\bibitem{SSN08}
L.~A.~Nunes Amaral, M.~Sales-Pardo, and M.~J. Stringer.
\newblock Effectiveness of journal ranking schemes as a tool for locating
  information.
\newblock {\em PLoS ONE}, 3(2):e1683, 2008.

\bibitem{JKR00}
K.~S. Jones, S.~Walker, and S.~E. Robertson.
\newblock A probabilistic model of information retrieval: Development and
  comparative experiments: Part 1.
\newblock {\em Information Processing \& Management}, 36(6):779--808, 2000.

\bibitem{B07}
C.~T. Bergstrom.
\newblock Eigenfactor: Measuring the value and prestige of scholarly journals.
\newblock {\em C\&RL News}, 68(5):314--316, 2007.

\bibitem{BWW08}
C.~T. Bergstrom, J.~D. West, and M.~A. Wiseman.
\newblock The {Eigenfactor} metrics.
\newblock {\em Journal of Neuroscience}, 28(45):11433--11434, 2008.

\bibitem{Eigenfactor}
J.~West, B.~Althouse, C.~Bergstrom, M.~Rosvall, and T.~Bergstrom.
\newblock {Eigenfactor.org -- Ranking and mapping scientific knowledge}.
\newblock Accessed July 1, 2009, at \texttt{http://www.eigenfactor.org}, 2009.

\bibitem{PN76}
G.~Pinski and F.~Narin.
\newblock Citation influence for journal aggregates of scientific publications:
  Theory, with application to the literature of physics.
\newblock {\em Information Processing \& Management}, 12(5):297 -- 312, 1976.

\bibitem{H65}
C.~H. Hubbell.
\newblock An input-output approach to clique identification.
\newblock {\em Sociometry}, 28:377--399, 1965.

\bibitem{L41}
W.~W. Leontief.
\newblock {\em The Structure of American Economy, 1919-1929}.
\newblock Harvard University Press, 1941.

\bibitem{BP98}
S.~Brin and L.~Page.
\newblock The anatomy of a large-scale hypertextual web search engine.
\newblock {\em Computer networks and ISDN systems}, 30(1-7):107--117, 1998.

\bibitem{GMLG01}
A.~A. Goodrum, K.~W. McCain, S.~Lawrence, and C.~L. Giles.
\newblock Scholarly publishing in the internet age: A citation analysis of
  computer science literature.
\newblock {\em Information Processing \& Management}, 37(5):661--675, 2001.

\bibitem{ZL02}
D.~Z. Zhao and E.~Logan.
\newblock Citation analysis using scientific publications on the web as data
  source: A case study in the xml research area.
\newblock {\em Scientometrics}, 54(3):449--472, 2002.

\bibitem{W02}
K.~M. Whitley.
\newblock Analysis of {SciFinder Scholar and Web of Science} citation searches.
\newblock {\em Journal of the American Society for Information Science and
  Technology}, 53(14):1210--1215, 2002.

\bibitem{BB05}
K.~Bauer and N.~Bakkalbasi.
\newblock An examination of citation counts in a new scholarly communication
  environment.
\newblock {\em D-Lib Magazine}, 11(9), 2005.
\newblock Retrieved December 20, 2008, from
  \texttt{http://www.dlib.org/dlib/september05/bauer/09bauer.html}.

\bibitem{J05}
P.~Jacs\`{o}.
\newblock As we may search. comparison of major features of the {Web of
  Science, Scopus, and Google Scholar} citation-based and citation-enhanced
  databases.
\newblock {\em Current Science}, 89(9):1537--1547, 2005.
\newblock Retrieved December 20, 2008, from
  \texttt{http://www.ias.ac.in/currsci/nov102005/1537.pdf}.

\bibitem{N05}
A.~Noruzi.
\newblock {Google Scholar}: The new generation of citation indexes.
\newblock {\em Libri}, 55(4):170--180, 2005.

\bibitem{PS05}
D.~Pauly and K.~I. Stergiou.
\newblock Equivalence of results from two citation analyses: {Thomson ISI}'s
  citation index and {Google}'s scholar service.
\newblock {\em Ethics in Science and Environmental Politics}, pages 33--35,
  2005.

\bibitem{BBGW06}
N.~Bakkalbasi, K.~Bauer, J.~Glover, and L.~Wang.
\newblock Three options for citation tracking: {Google Scholar, Scopus and Web
  of Science}.
\newblock {\em Biomedical Digital Libraries}, 7, 2006.
\newblock Retrieved December 20, 2008 from
  \texttt{http://www.pubmedcentral.nih.gov/articlerender.fcgi?artid=1533854}.

\bibitem{KT07}
K.~Kousha and M.~Thelwall.
\newblock Google scholar citations and {Google Web/URL} citations: A
  multi-discipline exploratory analysis.
\newblock {\em Journal of the American Society for Information Science and
  Technology}, 58(7):1055--1065, 2007.

\bibitem{S06}
G.~Saad.
\newblock Exploring the h-index at the author and journal levels using
  bibliometric data of productive consumer scholars and business-related
  journals respectively.
\newblock {\em Scientometrics}, 69(1):117--120, 2006.

\bibitem{NO07}
M.~Norris and C.~Oppenheim.
\newblock Comparing alternatives to the {Web of Science} for coverage of the
  social sciences literature.
\newblock {\em Journal of Informetrics}, 1(2):161--169, 2007.

\bibitem{BLL07}
J.~Bar-Ilan, M.~Levene, and A.~Lin.
\newblock Some measures for comparing citation databases.
\newblock {\em Journal of Informetrics}, 1(1):26--34, 2007.

\bibitem{MY07}
L.~I. Meho and K.~Yang.
\newblock Impact of data sources on citation counts and rankings of {LIS}
  faculty: {Web of Science vs. Scopus and Google Scholar}.
\newblock {\em Journal of the American Society for Information Science and
  Technology}, 58(13):2105--2125, 2007.

\bibitem{KT08}
K.~Kousha and M.~Thelwall.
\newblock Sources of {Google Scholar} citations outside the {Science Citation
  Index}: a comparison between four science disciplines.
\newblock {\em Scientometrics}, 74(2):273--294.

\bibitem{SV08}
D.~Shaw and L.~Vaughan.
\newblock A new look at evidence of scholarly citation in citation indexes and
  from web sources.
\newblock {\em Scientometrics}, 74(2):317--330, 2008.

\bibitem{MR08}
L.~I. Meho and Y~Rogers.
\newblock Citation counting, citation ranking, and h-index of human-computer
  interaction researchers: a comparison between {Scopus} and {Web of Science}.
\newblock {\em Journal of the American Society for Information Science and
  Technology}, 2008.

\bibitem{B08}
J.~Bar-Ilan.
\newblock Which h-index? a comparison of {WoS, Scopus and Google Scholar}.
\newblock {\em Scientometrics}, 74(2):257--271, 2008.

\bibitem{S08}
M.~Sanderson.
\newblock Revisiting h measured on {UK LIS} academics.
\newblock {\em Journal of the American Society for Information Science and
  Technology}, 59(7):1184--1190, 2008.

\bibitem{BMSRTD09}
L.~Bornmann, W.~Marx, H.~Schier, E.~Rahm, A.~Thor, and H.-D. Daniel.
\newblock Convergent validity of bibliometric {Google Scholar} data in the
  field of chemistry citation counts for papers that were accepted by
  {Angewandte Chemie International Edition} or rejected but published
  elsewhere, using {Google Scholar, Science Citation Index, Scopus, and
  Chemical Abstracts}.
\newblock {\em Journal of Informetrics}, 3(1):27--35, 2009.

\bibitem{P97}
V.~Pareto.
\newblock {\em Cours d'\'{e}conomie politique}, volume~2.
\newblock Universit\'{e} de Lausanne, Lausanne, 1897.

\bibitem{R98}
S.~Redner.
\newblock How popular is your paper? {An} empirical study of the citation
  distribution.
\newblock {\em The European Physical Journal B}, 4:131--134, 1998.

\bibitem{LS98}
J.~Laherr\'{e}re and D.~Sornette.
\newblock Stretched exponential distributions in nature and economy: ``fat''
  tails with characteristic scales.
\newblock {\em The European Physical Journal B}, 2:525--539, 1998.

\bibitem{S57}
W.~Shockley.
\newblock On the statistics of individual variations of productivity in
  research laboratories.
\newblock {\em Proceedings of the IRE}, 45:279--290, 1957.

\bibitem{P76}
D.~de~Solla~Price.
\newblock A general theory of bibliometric and other cumulative advantage
  processes.
\newblock {\em Journal of the American Society for Information Science},
  27:292--306, 1976.

\bibitem{M68}
R.~K. Merton.
\newblock The {Matthew Effect} in science.
\newblock {\em Science}, 159(3810):56--63, 1968.

\bibitem{CCTB83}
M.~Callon, J-P. Courtial, W.~Turner, and S.~Brain.
\newblock From translations to problematic networks: An introduction to co-word
  analysis.
\newblock {\em Social Science Information}, 22:191--235, 1983.

\bibitem{C94}
J-P. Courtial.
\newblock A co-word analysis of scientometrics.
\newblock {\em Scientometrics}, 31(3):251--260, 1994.

\bibitem{S73}
H.~Small.
\newblock Co-citation in the scientific literature: A new measure of the
  relationship between two documents.
\newblock {\em Journal of the American Society for Information Science and
  Technology}, 24:265--269, 1973.

\bibitem{K63}
M.~M. Kessler.
\newblock Bibliographic coupling between scientific papers.
\newblock {\em American Documentation}, 14:10--25, 1963.

\bibitem{SG83}
S.~K. Sen and S.~K. Gan.
\newblock A mathematical extension of the idea of bibliographic coupling and
  its applications.
\newblock {\em Annals of Library Science and Documentation}, 30:78--82.

\bibitem{GC96}
W.~Gl\"{a}nzel and H.~J. Czerwon.
\newblock A new methodological approach to bibliographic coupling and its
  application to the national, regional and institutional level.
\newblock {\em Scientometrics}, 37:195--221.

\bibitem{A73}
M.~R. Anderberg.
\newblock {\em Cluster analysis for applications}.
\newblock Academic Press, 1973.

\bibitem{JD88}
R.~C. Dubes and A.~K. Jain.
\newblock {\em Algorithms for clustering data}.
\newblock Prentice-Hall, Inc., 1988.

\bibitem{G85}
T.~F. Gonzalez.
\newblock Clustering to minimize the maximum intercluster distance.
\newblock {\em Theoretical computer science}, 38:293--306, 1985.

\bibitem{M67}
J.~B. MacQueen.
\newblock Some methods for classification and analysis of multivariate
  observations.
\newblock In {\em Berkeley symposium on mathematical statistics and
  probability}, pages 281--297, 1967.

\bibitem{J67}
S.~C. Johnson.
\newblock Hierarchical clustering schemes.
\newblock {\em Psycometrika}, 4:58--67, 1967.

\bibitem{GS76}
T.~F. Gonzalez and S.~Sahni.
\newblock P-complete approximation problems.
\newblock {\em Journal of the ACM}, 23:555--565, 1976.

\bibitem{P01}
K.~Pearson.
\newblock On lines and planes of closest fit to systems of points in space.
\newblock {\em Philosophical Magazine}, 2(6):559--572, 1901.

\bibitem{J02}
I.~Jolliffe.
\newblock {\em Principal component analysis}.
\newblock Springer, 2002.

\bibitem{RB07}
M.~Rosvall and C.~T. Bergstrom.
\newblock An information-theoretic framework for resolving community structure
  in complex networs.
\newblock {\em Proceedings of the National Academy of Sciences of USA},
  104(18):7327--7331, 2007.

\bibitem{RB08}
M.~Rosvall and C.~T. Bergstrom.
\newblock Maps of random walks on complex networks reveal community structure.
\newblock {\em Proceedings of the National Academy of Sciences of USA},
  105(4):1118--1123, 2008.

\bibitem{H52}
D.~Huffman.
\newblock A method for the construction of minimum-redundancy codes.
\newblock {\em Proceedings of the IRE}, 40:1098--1101.

\bibitem{S48}
C.E. Shannon.
\newblock A mathematical theory of communication.
\newblock {\em Bell System Technical Journal}, 27:379--423.

\end{thebibliography}

\end{document}